# Approaching the Intrinsic Photoluminescence Linewidth in Transition Metal Dichalcogenide Monolayers

*Obafunso A. Ajayi[1], Jenny V. Ardelean[1], Gabriella D. Shepard[2], Jue Wang[3], Abhinandan Antony[1], Takeshi Taniguchi[4], Kenji Watanabe[4], Tony F. Heinz[5], Stefan Strauf[2a], X.-Y. Zhu[3a], James C. Hone[1a]*

[1] Department of Mechanical Engineering, Columbia University, New York, NY 10027

[2] Department of Physics and Engineering Physics, Stevens Institute of Technology, Castle Point on the Hudson, Hoboken, NJ 07030

[3] Department of Chemistry, Columbia University, New York, NY 10027

[4] National Institute for Materials Science, 1-1 Namiki, Tsukuba, Japan

[5] Department Applied Physics, Stanford University, Stanford, CA 94305, USA and SLAC National Accelerator Laboratory, Menlo Park, CA 94025, USA

a) corresponding authors

ABSTRACT. **Excitonic states in monolayer transition metal dichalcogenides (TMDCs) have been the subject of extensive recent interest. Their intrinsic properties can, however, be obscured due to the influence of inhomogeneity in the external environment. Here we report methods for fabricating high quality TMDC monolayers with narrow photoluminescence (PL) linewidth approaching the intrinsic limit. We find that encapsulation in hexagonal boron nitride (h-BN) sharply reduces the PL linewidth, and that passivation of the oxide substrate by an alkyl monolayer further decreases the linewidth and also minimizes the charged exciton (trion) peak. The combination of these sample preparation methods results in much reduced spatial variation in the PL emission, with a full-width-at-half-maximum as low as 1.7 meV. Analysis of the PL line shape yields a homogeneous width of 1.43±0.08 meV and inhomogeneous broadening of 1.1±0.3 meV.**

Transition metal dichalcogenides (TMDCs) exhibit fascinating physical properties due to their two-dimensional (2D) nature. This is exemplified by their drastically reduced Coulomb screening and pronounced many-body interactions, leading to excitons[1–3] and trions[4] with respective binding energies of hundreds and tens of meV. Additionally, TMDCs show the unique physical properties of valley selectivity[5,6] and strong light-matter interactions,[7] which make these materials intriguing candidates for optical applications ranging from light detection and emission to quantum computing. Developing a full understanding of the photophysics of TMDC excitons and reaching the potential of TMDCs for applications will require producing materials with optical properties dominated by intrinsic factors, such as their radiative lifetime and electron-phonon interactions, rather than by extrinsic factors such as chemical, mechanical, electrostatic, or dielectric disorder. In high quality epitaxial quantum wells, a related low-dimensional system, the intrinsic PL linewidth has been observed,[8,9] but it is difficult to reach in nano-materials. For example, exciton emission from ultraclean carbon nanotubes can reach the intrinsic lifetime limit only if they are directly grown over an air gap, with little interference from the local chemical and physical environment.[10] In the case of TMDC monolayers, the extreme sensitivity of atomically thin materials to environmental effects and external disorder[11] has hindered observation of photophysical properties in the intrinsic regime.

The photoluminescence (PL) linewidth observed for exciton emission is a crucial indicator of material quality, and reflects intrinsic contributions from the radiative lifetime and dephasing from exciton-phonon interactions, as well as extrinsic inhomogeneous broadening effects from factors such as defects and substrate-induced disorder. For TMDC monolayers supported on $SiO_2$ substrates, the PL linewidth is typically exceeds 10 meV at low temperatures. Recent studies. employing four-wave mixing techniques indicate that this linewidth is dominated by inhomogeneous broadening, with an underlying homogeneous contribution of about 2 meV.[12–16] Likewise, the measured exciton emission time has exhibited a large variation associated with significant nonradiative decay channels.[12,17–19] These studies provide strong motivation to identify methods of sample preparation through which the intrinsic PL linewidth can be reached. An additional important parameter indicative of sample quality in TMDCs is the ratio between the neutral exciton and trion intensity, which has been seen to be strongly dependent on carrier density: the neutral exciton dominates at low density and the trion at high density.[20] It is not presently clear to what degree the carrier density in $SiO_2$-supported TMDCs results from intrinsic doping due to vacancies and impurities vs. from extrinsic doping (from substrate charge traps).

In this work we have systematically investigated the effects of h-BN encapsulation and $SiO_2$ surface passivation on important features - linewidth, trion/neutral exciton ratio (indicative of doping), and lateral spatial inhomogeneity - of the PL spectrum of a model TMDC, monolayer MoSe$_2$. Hexagonal boron nitride (h-BN) has proven to be a near-ideal substrate for 2D materials such as graphene and TMDCs. By isolating the materials from the charge disorder at the surface of SiO$_2$, h-BN encapsulation has been demonstrated to reduce doping and charge inhomogeneity, thus dramatically increasing electron mobility.[21,22] As a consequence, h-BN encapsulation has become standard practice in the fabrication of high-quality 2D devices for electronic applications. Substrate surface passivation by self-assembled monolayers has also been demonstrated as an effective means to increase electron mobility and realize previously elusive *n*-type doping in organic semiconductor FET devices.[23] This was achieved by SiO$_2$ surface passivation with an organic dielectric monolayer that eliminated hydroxyl groups, and thus electron traps, from the SiO$_2$ surface. Despite the benefits of surface passivation and h-BN encapsulation in electronic applications, these practices have not yet been widely adopted or investigated in optical studies of TMDCs.

In the present work, we fabricated and investigated four types of samples: i) monolayer MoSe$_2$ on an untreated SiO$_2$/Si substrate (M/SiO$_2$); ii) monolayer MoSe$_2$ on a passivated SiO$_2$/Si substrate (M/P-SiO$_2$); iii) monolayer MoSe$_2$ doubly encapsulated with h-BN on an untreated SiO$_2$/Si substrate (BMB/SiO$_2$); and iv) monolayer MoSe$_2$ doubly encapsulated with h-BN on a passivated SiO$_2$/Si substrate (BMB/P-SiO$_2$). We mechanically exfoliated MoSe$_2$ monolayers from a commercially obtained bulk crystal (HQ Graphene) onto an SiO$_2$/Si substrate (with a 285 nm thick SiO$_2$ layer) using a combination of oxygen plasma treatment and heating that has been shown[24] to produce larger monolayer flakes. We also mechanically exfoliated high-quality h-BN layers (without heating) and characterized them with atomic force microscopy to identify flakes 20-30 nm thick with clean and ultra-flat areas 20 ~ 25 μm in size, an important requirement for minimizing sample disorder. Encapsulated h-BN/MoSe$_2$/h-BN samples were prepared using a dry transfer method with a PC/PDMS lens (PC- poly(bisphenol A carbonate) & PDMS - poly(dimethyl siloxane)) for sequential pickup of the flakes. The pickup process was carried out slowly, with heating/cooling at a rate of ~ 0.5°C/min and a vertical translation rate of ~0.25 μm/s to minimize cracks and trapped bubbles at the h-BN/MoSe$_2$ interface that could lead to anomalous signatures in our TMDC's PL features. Passivated SiO$_2$/Si substrates were prepared as follows: the substrates were cleaned by rinsing with CH$_2$Cl$_2$, dried, and then oxidized in 3:1 conc. H$_2$SO$_4$: 30% H$_2$O$_2$ for 2 hours at 100° C, followed by thorough rinsing with DI H$_2$O. (Caution: the so-called piranha solution is a strong oxidant and should be handled with care.) The self-assembled monolayer (SAM) coating on cleaned SiO$_2$ surface was obtained from MicroSurfaces, Inc. (http://microsurfacesinc.com); the hydrophobic surface is characterized by a water contact angle of ~100$^o$.

We carried out PL measurements at a temperature of 3.8 K using a closed-cycle He cryostat system with piezo-scanner for mapping (Attocube Attodry 1100). For the PL measurements, the samples were illuminated at a wavelength of 532 nm at low power with a cw diode laser (~5 nW for PL intensity mapping and typically < 4μW for the PL spectra). The spectra were captured on a liquid nitrogen cooled silicon CCD camera after being dispersed in a grating spectrometer with 70 μeV resolution at the relevant wavelengths. For intensity mapping, the spectral emission was coupled into a multimode fiber and sent to a silicon single photon counting avalanche photodiode.

In **Figure 1** we present typical PL spectra for the four different types of sample. Each spectrum shows two peaks, corresponding to the neutral exciton (the higher energy peak) and trion (the lower energy peak).[20] For the bare monolayer on an untreated substrate (M/SiO$_2$), the peaks have a full-width at half maximum (FWHM) linewidth of ~9 meV. The trion peak shows ~50% greater intensity than the exciton, indicating substantial static doping (**Fig. 1a**). In addition, a low-energy 'tails' extends more than 50 meV below the trion line. h-BN encapsulation (BMB/SiO₂, **Fig. 1b**) narrows both exciton and trion peaks substantially (to 3 meV), with a clear decrease in the low-energy tail. SiO₂ surface passivation (M/P-SiO₂, **Fig. 1c**)  narrows the peak width to a lesser degree (to 6 meV), but dramatically reduces the trion intensity. Finally, combining both treatments (BMB/P-SiO₂, Fig. 1d) results in the lowest trion intensity and narrowest emission line (2 meV). We also observe that the exciton emission energy from BNB/P-SiO₂ in Fig. 1d is red-shifted by about 10 meV compared to other samples, likely due to a reduction in strain[25–27] of the BMB stack on the hydrophobic treated SiO₂ substrate. These findings suggest that the dominant effect of h-BN-encapsulation is to reduce the charge disorder by spatially separating the TMDC layer from the charged SiO₂ surface, giving rise to a significantly narrower linewidth. Similarly, passivation of the SiO₂ improves the emission characteristics predominantly by reducing static charging. The combination of h-BN-encapsulation and $SiO_2$ surface passivation yields the narrowest PL linewidth for the $WSe_2$ monolayer, with an emission width approaching the intrinsic limit,[12–16] and also drastically reduces doping from the extrinsic electrostatic landscape. The remaining trion intensity (Fig. 1d) which is more than one order of magnitude lower than that on the bare $SiO_2$ surface (Fig. 1a) may be attributed to intrinsic doping from structural defects in the $WSe_2$ monolayer.

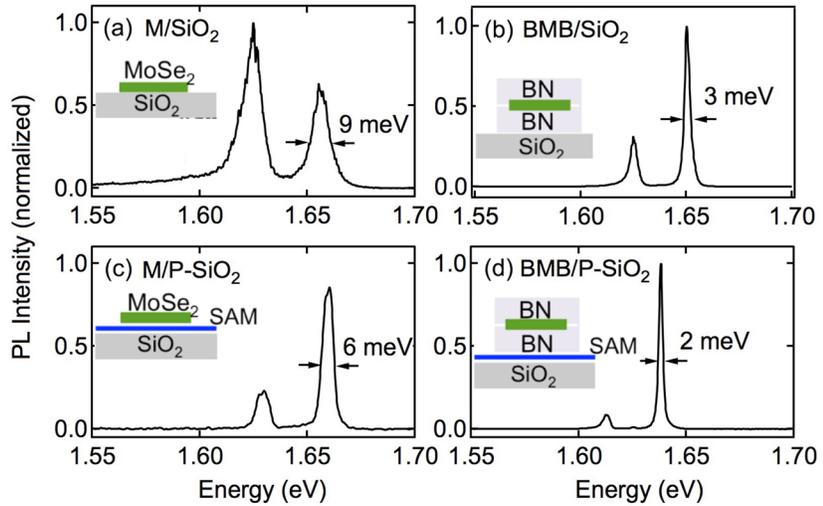

**Figure 1.** Representative PL spectra of MoSe₂ monolayer at 4K: (**a**) Monolayer on an untreated SiO₂ substrate (M/SiO₂); (**b**) h-BN-encapsulated monolayer on an untreated SiO₂ substrate (BMB/SiO₂); (**c**) monolayer on a passivated SiO₂ (M/SiO₂) substrate; (**d**) h-BN-encapsulated monolayer on a passivated SiO₂ substrate. (BMB/SiO₂).

In order to study the lateral spatial variation of these features across the sample, we recorded hyperspectral images of the exciton and trion features. This was accomplished by sending the PL through narrow bandpass filters (FHWM=10 nm) centered at the respective peaks at 750 nm (1.653 eV) and 765 nm (1.620 eV) before reaching the detector. In **Figures 2a,b** we present normalized exciton and trion spatial intensity maps for a representative M/SiO₂ sample,. We observe comparable exciton/trion intensities over the entire sample, with substantial intensity variation on the scale of a few μm. For a BMB/P-SiO₂ sample, we see (**Figures 2d,e**) reduced intensity variation and trion intensity, except for a prominent bright trion spot (with a corresponding dark spot in the exciton intensity) that results from a gold alignment marker on the surface. **Figs. 2c and 2f** show spectra taken at the three indicated locations on each sample. For the M/SiO₂ sample, we observe fluctuations of about 5 meV in the position of the emission peaks, whereas the BMB/P-SiO₂ sample exhibits no measurable variation in peak position. The latter will be addressed in more detail in Fig. 4. In the following, we present more quantitative data on the spatial variation in the emission characteristics of the different samples.

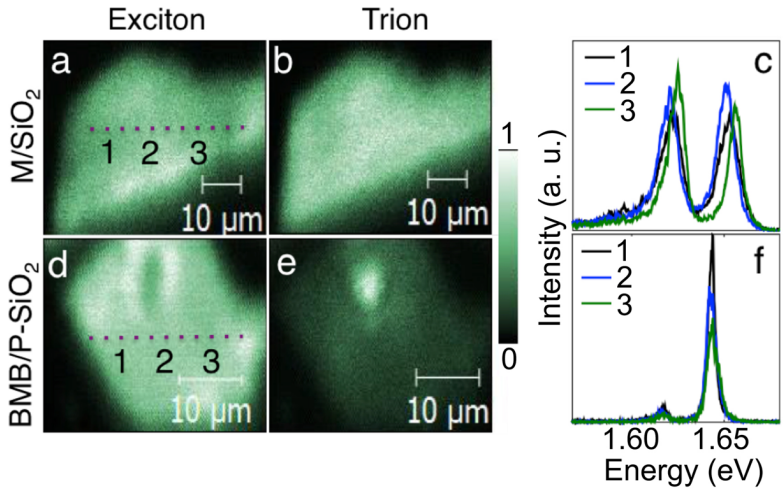

**Figure 2.** Spatial intensity mapping of exciton and trion emission and spectral linescans for a $MoSe_2$ monolayer. (**a**) Exciton spatial map for a monolayer on an untreated $SiO_2$ substrate (M/$SiO_2$). (**b**) Trion spatial map for a monolayer on an untreated $SiO_2$ substrate (M/$SiO_2$). (**c**) PL spectra at the positions indicated in (a) for the M/$SiO_2$ sample. (**d**) Exciton spatial map of an h-BN-encapsulated monolayer on a passivated $SiO_2$ substrate (BMB/P-$SiO_2$). (**e**) Trion spatial map of an h-BN-encapsulated monolayer on a passivated $SiO_2$ substrate (BMB/P-$SiO_2$) (**f**) PL spectra at positions indicated in (d) for the BMB/P-$SiO_2$ sample.

In **Figure 3**, we display histograms of total PL intensity (a,c) and exciton/trion ratio (b,d) from all the pixels in the PL images in the wrinkle/tear-free locations away from the edges. Both SiO₂ surface passivation and encapsulation dramatically reduce the intensity variation, with the standard deviation dropping by from 41±10% on M/SiO₂ to 19±2% on M/P-SiO₂, BMB/SiO₂, or BMB/P-SiO$_2$,. A particularly significant change is in the trion/exciton ratio with the passivation of the SiO$_2$ surface. This 1.2±.3 for M/SiO$_2$ and 0.7±0.3 for BMB/SiO2 to 0.4±0.2 for M/P-SiO$_2$ and 0.18±0.06 for BMB/P-SiO$_2$.

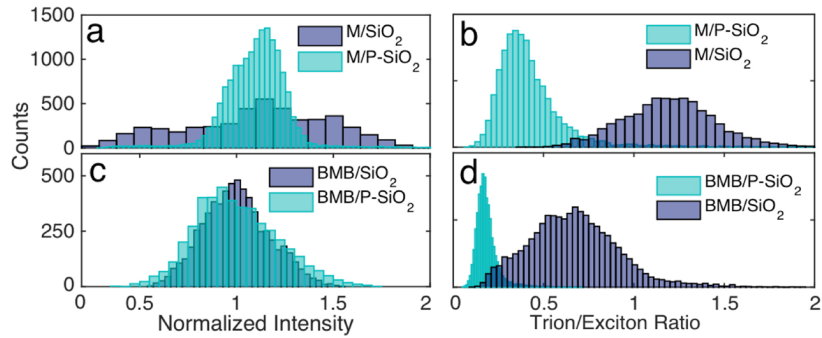

**Figure 3.** Normalized intensity and trion/exciton ratio histograms for all sample types: (**a**) Normalized intensity distributions for monolayer samples on untreated (M/SiO₂) and passivated (M/P-SiO₂) substrates. (**b**) Trion/exciton ratio distribution for monolayer samples (M/SiO₂ and M/P-SiO₂). (**c-d**) As in (a-b) but for h-BN encapsulated substrates on untreated and passivated substrates (BMB/SiO₂, BMB/P-SiO₂).

These results establish that h-BN-encapsulation and SiO$_2$ surface passivation both reduce broadening and the trion/exciton ratio in PL spectra. The best case scenario is obtained when we combine the two approaches and resulting exciton PL linewidth (2 meV) comparable to the homogeneous limit obtained in four-wave mixing experiments.[12–16] In the following, we carry out a quantitative analysis of the exciton PL peaks for BMB/P-SiO$_2$ samples.

**Figure 4a** shows representative PL spectra in the region of the exciton peak from 10 random locations on a BMB/P-SiO$_2$ sample. We plot the spectra on a logarithmic intensity scale to reveal weak features in the wings of the spectra. Firstly, we observe variation in the peak exciton energy from different positions. We find that six of the ten spectra are characterized by excitonic peak positions of $E_{ex}$ = 1.6391±0.0001 eV, four are red-shifted by as much as 0.003 eV (*e.g.*, the red spectrum in **Figure 4a**). Thus we conclude that, while the majority of locations on the BMB/P-SiO$_2$ sample show homogeneity in PL emission energy, there are also locations with red-shifts in the excitonic transition; the latter likely result from variations local strain.[25–27] Secondly, we note the asymmetry in the peak shape for intensities over one-order-of-magnitude lower than that of the main exciton transition. The intensity is enhanced on the low energy side, which may reflect phonon side bands of PL emission. Thirdly, there is an additional high-energy shoulder, with an intensity reduced more than one order of magnitude lower and peak positions varying from spectrum to spectrum. The origin of this high-energy shoulder is not known, but the blue-shifted position suggests dielectric disorder as the possible explanation. The presence of a local "bubble" would decrease the effective dielectric screening, leading to a blue shift in the exciton

energy.[28,29]

**Figures 4b,c** show two representative PL spectrum from the group of six spectra with $E_{ex}$ = 1.6391±0.0001 eV in **Figure 4a** presented on a linear scale. **Figures 4b** is a spectrum with a clearly visible high-energy shoulder, while this feature is nearly invisible in the spectrum in **Figures 4c**. The intrinsic line-shape from exciton emission is commonly assumed to be Lorentzian, but this may be modified in a TMDC monolayer due to exciton-phonon scattering at non-zero temperatures, exciton-charge scattering in the presence of finite doing, and exciton relaxation to lower energy states such as spin-forbidden dark excitons. While how these issues deserve further quantitative studies, here adopt a first-order approximation and assume that each narrow PL peak comes from an intrinsic

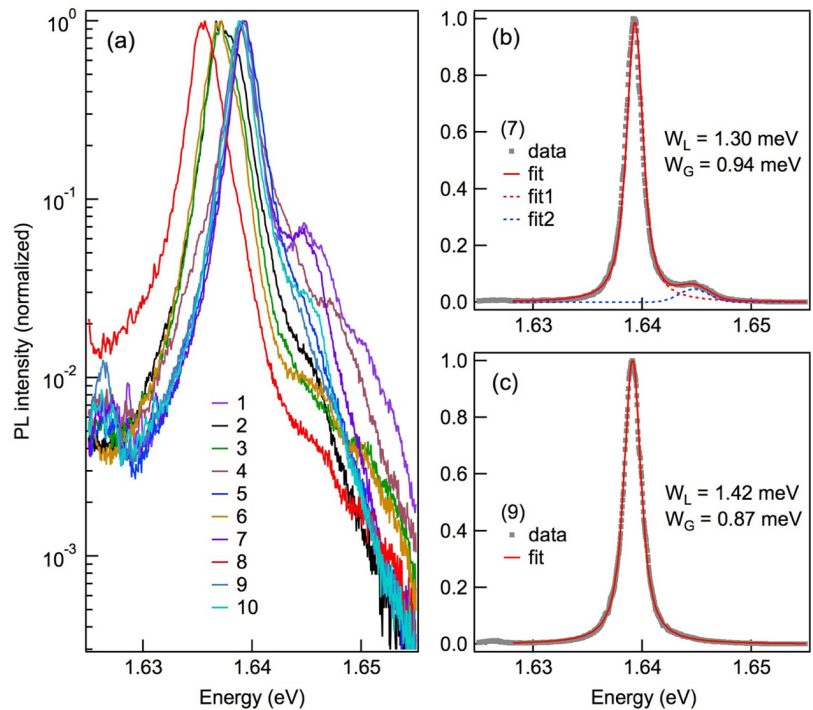

**Figure 4.** Spectra and lineshape analysis for the neutral exciton for a BMB / p-SiO$_2$ sample. (**a**) Exciton peak for ten points chosen at random, with the intensity on a logarithmic scale. (**b,c**) Spectra at points 7, 9 from (a), shown on a linear scale. Both peaks are fit well by a Voigt function, with similar values of the homogenous linewidth $W_L$ and inhomogeneous broadening $W_G$ as shown. The high-energy shoulder present in both spectra and seen clearly in (**b**) is fit by an additional Gaussian peak.

Lorentian shape broadened by a Gaussian distribution due to inhomogeneity. We fit the main exciton peak with a Voigt function, *i.e.,* a convolution of a Lorentzian and a Gaussian, to represent the influence of homogeneous and inhomogeneous broadening. We add a Gaussian feature (blue-dotted spectrum) to the fit to represent the secondary peak. The fits yield FWHMs of the Lorentzian ($W_L$) and the Gaussian ($W_L$) of the main exciton peaks, as shown in **Figures 4b,c**. From fits to all ten spectra, we obtain $W_L = 1.43\pm0.08$ meV and $W_G = 1.1\pm0.3$ meV. The average value of the total FWHM is $2.0\pm0.2$ meV, with the narrowest line displaying FWHM = $1.7\pm0.1$ meV.

We summarize in **Table 1** the measured FHWM, along with the Lorentzian ($W_L$) and Gaussian ($W_L$) width from the fits for samples prepared according to the different protocols. For the most favorable case of h-BN encapsulated samples on the passivated substrates (BMB/P-$SiO_2$), we consistently find an inhomogeneous linewidth ($W_G$) smaller than the homogeneous width ($W_L$), indicating that the PL spectrum from the combined h-BN-encapsulation and $SiO_2$ surface passivation method is approaching the intrinsic limit.

**Table 1**. Summary of homogeneous and inhomogeneous broadening in all sample types.

| Sample Type | Total emission linewidth (FWHM, meV) | Lorentzian width (meV) | Gaussian width (meV) |
|---|---|---|---|
| M/$SiO_2$ | $9.8\pm2.8$ | $3.7\pm0.7$ | $6.5\pm0.9$ |
| M/P-$SiO_2$ | $5.7\pm1.3$ | $2.9\pm0.5$ | $3.2\pm1.3$ |
| BMB/$SiO_2$ | $3.5\pm1.0$ | $1.7\pm0.5$ | $2.5\pm0.9$ |
| BMB/P-$SiO_2$ | $2.0\pm0.2$ | $1.43\pm0.08$ | $1.1\pm0.3$ |

In summary, we show that the use of h-BN encapsulation and substrate surface passivation drastically reduces disorder in $WSe_2$ monolayers, leading to an 80% reduction in the PL linewidth from that of monolayer on the untreated $SiO_2$ layer. The PL linewidth for monolayer TMDC on the commonly used $SiO_2$ surface is dominated by heterogeneity and disorder. We also show that $SiO_2$ surface passivation and h-BN-encapsulation effectively reduce electron doping of the $WSe_2$ monolayer by more than one order of magnitude based on the decrease in the strength of trion emission. By strongly reducing local disorder in the substrate, we demonstrate a very significant reduction in the inhomogeneous linewidth. We have achieved an emission linewidth 2.0±0.2 meV, which is predominantly determined by the homogeneous linewidth of 1.43±0.08 meV according to our line shape analysis. During the writing of this manuscript, we became aware of work reporting a similar PL linewidth reduction for $MoS_2$ monolayers encapsulated by hBN.[30] The realization of samples with such spectrally narrow spectral features should assist in emerging applications of these materials in optoelectronics and valleytronics, as well as in fundamental studies of exciton-charge, exciton-phonon, and exciton-exciton interactions, as well as in interlayer charge transfer exciton formation at TMDC heterojunctions[31].

**ACKNOWLEDGMENT.** This manuscript is primarily based upon work supported by the NSF MRSEC program through Columbia in the Center for Precision Assembly of Superstratic and Superatomic Solids (DMR-1420634). XYZ acknowledges support from NSF DMR 1608437 for analysis in Figure 4. S.S. acknowledges support by the National Science Foundation (NSF) under award DMR-1506711, and financial support for the attodry1100 system under NSF award ECCS-MRI-1531237.